\documentclass[twocolumn,aps,pra,showpacs]{revtex4}            
\usepackage{graphics,dcolumn}
\usepackage{bm}
\usepackage[fleqn]{amsmath}
\usepackage{hyperref}
\hypersetup{
     colorlinks=true,      
    linkcolor=blue,        
    citecolor=blue,        
    filecolor=magenta, 
    urlcolor=cyan          
}
\usepackage{exscale,relsize}
\usepackage{epsfig}
\usepackage{amssymb}
\usepackage{amsmath}
\usepackage{euscript}
\usepackage{float}
\usepackage{tikz}
\usetikzlibrary{snakes}
\usepackage{ulem}
\usetikzlibrary{decorations.pathmorphing}
\usetikzlibrary{patterns}
\usepackage{bbm}
\usepackage{footmisc}

\newcommand{\expec}[1]{\langle #1\rangle}

\begin{document}

\date{\today}
\title{Coherent inelastic backscattering of laser light from three isotropic atoms}

\author{Andreas Ketterer} \email{andreas.ketterer@univ-paris-diderot.fr}\altaffiliation[Present address: Laboratoire Mat\'eriaux et Ph\'enom\`enes Quantiques, Universit\'e Paris Diderot, 10 rue A. Domon et L. Duquet, 75013 Paris, France]{}
\affiliation{Physikalisches Institut, Albert-Ludwigs-Universit\"at Freiburg, Hermann-Herder-Str. 3,
D-79104 Freiburg, Germany}
\author{Andreas Buchleitner}
\author{Vyacheslav N. Shatokhin}
\affiliation{Physikalisches Institut, Albert-Ludwigs-Universit\"at Freiburg, Hermann-Herder-Str. 3,
D-79104 Freiburg, Germany}
\begin{abstract}
We study the impact of double and triple scattering contributions on coherent
backscattering of laser light from saturated isotropic atoms, in the
helicity preserving polarization channel. Using the recently proposed
diagrammatic pump-probe approach, we analytically derive single-atom spectral responses to a classical polychromatic driving field, combine them self-consistently to double and triple scattering processes, and numerically deduce the corresponding elastic and inelastic spectra, as well as the total backscattered intensities. We find that account of the
triple scattering contribution leads to a faster decay of
phase-coherence with increasing saturation of the atomic transition
as compared to double scattering alone, and to a better agreement
with the experiment on strontium atoms.

\end{abstract}


\maketitle

\section{Introduction}

Coherent transport of light in a disordered medium can
effectively be investigated using a multiple scattering setting
where a laser field is injected into a dilute cloud of cold
atomic scatterers \cite{PhysRevLett.83.5266,PhysRevLett.88.203902}.
Constructive interference of counter-propagating
multiply scattered waves leads to coherent
backscattering (CBS) -- the enhancement of the average scattered
light intensity in backward direction \cite{akkermansMeso}. The degeneracy of the atomic dipole transitions and/or the nonlinear atomic response to an intense driving
field destroys perfect phase coherence of the interfering multiply scattered waves, and, consequently, causes a decrease of the CBS enhancement, as reported in recent
experiments \cite{PhysRevLett.83.5266,PhysRevE.70.036602,balik05}.

The development of a multiple scattering theory which accounts for 
the electronic structure of the atoms, is to this day a subject of active investigations \cite{kupriyanov06,mueller2011}. Besides CBS of light, progress in this field is crucial to assess the possibility to achieve Anderson localization of light \cite{akkermans08,skipetrov14}, or for the realization of random lasers \cite{savels07,guerin10,Baudouin:2013yg} with cold atoms.

In the linear scattering regime, transport theories based on diagrammatic scattering approaches \cite{PhysRevA.64.053804,0295-5075-61-3-327,kupriyanov03} yield excellent agreement with the experimental observations on rubidium (Rb) \cite{PhysRevLett.83.5266} and strontium (Sr)  \cite{PhysRevLett.88.203902} atoms. For example, these theories show that the reduction of the CBS signal for Rb atoms occurs due to their ground state degeneracy. Unfortunately, it is very difficult to generalize those diagrammatic theories to treat nonlinear inelastic scattering from saturated atoms. So far, a nonlinear transport theory in the atomic medium has been developed for two incident photons \cite{PhysRevA.73.013802}, which is far below the saturation regime probed in the experiments \cite{PhysRevE.70.036602,balik05}.

Quantum optical master equations and related approaches \cite{agarwal74,scully,cohen1998atom} are powerful tools for describing the nonlinear response of individual atoms to an intense laser field. These methods have also been used to characterize a double scattering contribution to CBS from two saturated, dipole-dipole interacting, randomly located atoms \cite{PhysRevLett.94.043603,PhysRevA.73.063813,gremaud06,PhysRevA.76.043832}. However, the exponential scaling of the atomic Hilbert space dimension with the number of scatterers precludes the application of the standard quantum optical methods to treat CBS off a dilute cloud of cold saturated atoms. 

Recently, a novel method was put forward \cite{Geiger2010244,PhysRevA.82.013832} to deal with the above problems.
In this framework, the nonlinear response of the atoms to the laser driving (pump) is accounted for nonperturbatively, while the response to the weak coherent fields (probes) scattered from the surrounding atoms is incorporated perturbatively, within {\it single}-atom optical Bloch equations (OBE) under {\it classical} polychromatic driving. Thereafter, the spectral response functions resulting from solutions of the ``polychromatic'' OBE serve as building blocks for a self-consistent, diagrammatic construction of the multiple scattering CBS signal. We will refer to this new method as the {\it diagrammatic pump-probe} (DPP) approach.

Since all the building blocks are essentially single-atom quantities, the problem of the exponential growth of the Hilbert space dimension is circumvented in the framework of the DPP approach.  Therein, multiple scattering of light in ensembles of saturated atoms is rendered into a form which befits Monte Carlo simulations \cite{binninger12}. Therefore, it is a promising method for the quantitative modelling of CBS of light off the bulk atomic medium. Moreover, for double scattering from two-level (scalar) atoms, the DPP equations yield analytical solutions that are strictly equivalent to those following from the two-atom master equation \cite{Shatokhin2010150}. The analytical equivalence between the two methods holds also in the case of triple scattering, provided that the terms responsible for recurrent scattering be dropped from the solutions of the three-atom master equation \cite{SlavaTriple}.

The DPP method has recently been generalized to vector atoms \cite{ralf,ralf13}. The numerical equivalence between the double scattering spectra, obtained within the DPP, and the master equation \cite{PhysRevA.76.043832} approaches, respectively, has been established for (isotropic) strontium atoms \cite{ralf}. By unifying the results of \cite{SlavaTriple} and \cite{ralf13}, it has become possible to derive arbitrary single-atom responses for atoms with arbitrary internal degeneracy \cite{Kneddi}. In particular, the single-atom spectral response functions needed for the precise calculation of {\it triple} scattering from saturated Sr atoms, have thus become available.

The assessment of triple scattering from saturated vector atoms is a challenging problem which has not been so far studied. Besides, such a study
provides an opportunity to improve on an earlier theoretical description \cite{PhysRevLett.94.043603} of the experiment \cite{PhysRevE.70.036602}.
Indeed, in the `saturation' experiment, with an optically thin cloud of Sr atoms \cite{PhysRevE.70.036602}, lowest-order multiple scattering sequences gave the main contribution to the CBS signal. Although a theoretical model based on double scattering \cite{PhysRevLett.94.043603} yields qualitatively correct results, they deviate quantitatively from the experimental ones.

In the present contribution, we use the DPP approach to calculate triple scattering CBS spectra from three isotropic atoms in the helicity preserving (h $\parallel$ h) polarization channel. In passing, we also present the general expressions for arbitrary elastic and inelastic spectral responses. These results can be used in future simulations of radiation transport in cold atoms with degenerate transitions. Using the obtained triple scattering spectra, as well as the already available results for double scattering \cite{PhysRevA.76.043832,ralf13}, we deduce
the CBS enhancement factor as a function of the saturation parameter. To this end, we combine double and triple scattering contributions, with phenomenologically adjusted relative weights, into a total signal. We show that the account of triple scattering leads to a faster decay of the CBS enhancement with the saturation parameter, than when only double scattering is included. Thereby, we attain a better agreement with the experiment \cite{PhysRevE.70.036602}.

The structure of this paper is as follows. In the next section, we
introduce the three-atom CBS model, and
outline how the double and triple scattering spectral signals from Sr atoms can be
calculated using the DPP approach. 
Section \ref{sec:num_results} presents our numerical results, such as triple scattering elastic and inelastic spectra, as well as the CBS enhancement factor,
as a function of the atomic saturation parameter. Finally, we conclude in Sec. \ref{sec:conclusion}.
\section{Diagrammatic pump-probe approach}
\label{sec:ThreeSrAtoms}
\subsection{Model}
\label{sec:model}
\begin{figure}
\begin{center}
\includegraphics[width=0.45\textwidth]{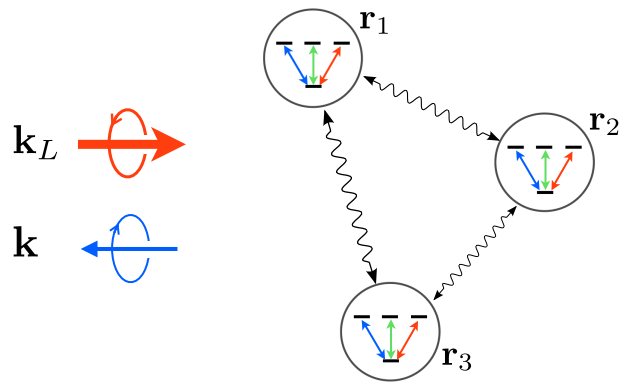}
\end{center}
\caption{(color online) Toy CBS model: three randomly located atoms at positions $\textbf{
r}_\alpha$ are driven by a circularly polarized laser field (red
arrow) with wave vector $\textbf k_L$. The atoms exchange photons
via the radiative dipole-dipole interaction (double wavy arrows).
The backscattered far-field with wave vector $\textbf k=-\textbf
k_L$ (blue arrow) is observed in the helicity preserving
polarization channel h$\parallel$h (that is, with flipped
polarization). Whereas the CBS signal originates from one transition
of the atoms (indicated by double blue arrows), all three atomic
dipole transitions are involved in the triple scattering process (see
text).} \label{fig:ThreeSrAtoms}
\end{figure}
Let us consider a toy CBS model depicted in Fig. \ref{fig:ThreeSrAtoms}.
Three immobile atoms, randomly located in free space, are driven by a circularly polarized, near-resonant, continuous wave laser
field with the amplitude ${\cal E}_L$, wave vector $\textbf k_L$, and frequency $\omega_L$.  We will focus on the average, stationary, backscattered light (spectral) intensity in the far-field, along the wave vector $\textbf k=-\textbf k_L$, in the parallel helicity polarization channel (h $\parallel$ h), that is, with flipped polarization.

To account for the waves' polarizations, we use indices $q=\pm 1,0$, which define the unit vectors $\hat{\bf e}_q$ in the spherical basis:
\begin{equation}
    \hat{\textbf{e}}_{\pm1}=\mp \frac{1}{\sqrt{2}} (\hat{\textbf{e}}_x \pm i \hat{\textbf{e}}_y), \quad \hat{\textbf{e}}_0=\hat{\textbf{e}}_z,
     \label{eqn:SphericalBasis}
     \end{equation}
where $\hat{\textbf e}_x$, $\hat{\textbf e}_y$ and $\hat{\textbf e}_z$ are the Cartesian unit vectors. Hence, the laser wave coming in along the $z$-axis, with circular polarization $\hat{\bf e}_L=\hat{\bf e}_{+1}$, is characterized by the index $+1$ and the detected wave by the index $-1$.

We consider atoms with the dipole transition $J_g=0\leftrightarrow J_e=1$, where $J$ is the total angular momentum of the ground ($g$) and excited ($e$) states, with transition frequency $\omega_0$. This isotropic transition, which corresponds to that of Sr atoms probed in the  experiment \cite{PhysRevE.70.036602}, is characterized by equal Clebsch-Gordan coefficients associated with the three components of the vector dipole operator, see Eq.~(\ref{dplus_general}) below.  Furthermore, each of the dipole transitions is characterized by the reduced matrix element $d$ and the line width of the excited state sublevel $2\gamma$.

Throughout this work, we assume that the atoms are in the far-field of each other, that is, $k_Lr_{\alpha\beta}\gg1$ (dilute regime), where $r_{\alpha\beta}=|{\bf r}_\alpha-{\bf r}_\beta|$ is the distance between atoms $\alpha$ and $\beta$ (see Fig.~\ref{fig:ThreeSrAtoms}). Therein, the smallness of the radiative dipole-dipole interaction constant, which scales as $(k_Lr_{\alpha\beta})^{-1}$, ensures that
each pair of atoms exchanges no more than a single photon. Under this condition, the multiple scattering signal from saturated atoms can be expressed via a self-consistent combination of single-atom building blocks \cite{PhysRevA.82.013832}.

In the h $\parallel$ h channel, single scattering is filtered out, and, within our model, the main contribution to the detected signal  originates from double scattering processes, including pairs of atoms in Fig.~\ref{fig:ThreeSrAtoms}. On top of that, there is a small correction due to triple scattering, where all atoms in Fig.~\ref{fig:ThreeSrAtoms} are involved. In contrast, in an atomic cloud, with a huge number ($\sim 10^8$) of scatterers, double and triple scattering can have contributions comparable in magnitude. This is precisely the situation we will mimic with our toy model. To this end, we first evaluate the double and triple scattering signals using the DPP approach. Thereafter, we combine them into the total signal with statistical weights that are adjusted to the optical thickness of the Sr cloud in \cite{PhysRevE.70.036602}. We describe a Monte Carlo simulation procedure, which we use to obtain the statistical weights, in Sec.~\ref{sec:DTweights}.

We now proceed with a presentation of the basic elements of the DPP approach, on the example of double scattering.

\subsection{Double scattering}
\label{sec:double}
\begin{figure}
\begin{center}
\includegraphics[width=0.49\textwidth]{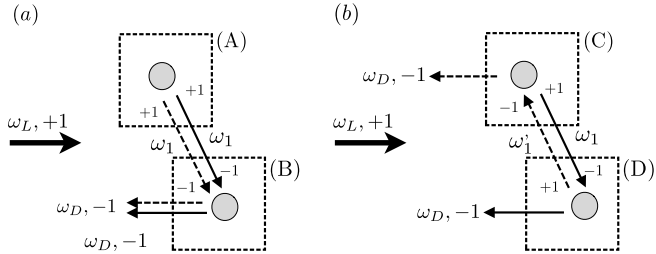}
\end{center}
\caption{Double scattering processes surviving the disorder average: (a) background (ladder) contribution consisting of co-propagating fields; (b) interference (crossed) contribution resulting from the interference of counter-propagating fields. Thick arrows depict the incoming laser wave with the frequency $\omega_L$. Thin solid and dashed arrows depict scattered positive- and negative-frequency fields, respectively. In general, the scattering processes are inelastic ($\omega_1\neq\omega_1^\prime\neq\omega_D\neq \omega_L$). The numbers $\pm 1$ encode the waves' polarizations, and correspond to the helicity-preserving channel (see explanation in the text). Both ladder and crossed contributions can be decomposed into pairs of single-atom blocks which are enclosed in dotted frames: (a) (A) and (B); (b) (C) and (D).}
\label{fig:DoubleScatCont}
\end{figure}

Let us consider either pair of atoms from Fig.~\ref{fig:ThreeSrAtoms}. There are two generic processes surviving the disorder average [see Fig.~\ref{fig:DoubleScatCont}]. The first process, composed of the co-propagating positive- and negative-frequency amplitudes  (here and henceforth, depicted by solid and dashed arrows, respectively), see Fig.~\ref{fig:DoubleScatCont}(a),
contributes to the background, or ladder, spectral intensity at the frequency $\omega_D$. The second process describes the interference between the counter-propagating amplitudes, and contributes to the so-called crossed spectral intensity, see Fig.~\ref{fig:DoubleScatCont}(b). 

The main idea of the DPP approach is to express the double (in general, multiple) scattering, stationary, spectrally-resolved CBS signal using single-atom spectral responses or building blocks \cite{PhysRevA.82.013832,Geiger2010244}. In Fig.~\ref{fig:DoubleScatCont}, we can select four single-atom blocks, enclosed in dotted frames: (A), (B), (C), and (D). The latter can be evaluated by solving the single-atom OBE under classical bichromatic (in general, polychromatic) driving, see Sec.~\ref{sec:generalOBE}. 

To write down the OBE, we need to specify the polarizations and frequencies of the fields with which the atom interacts, for each of the building blocks. Besides a laser-driven atom (represented by a gray circle), each of the building blocks (A)-(D) in Fig.~\ref{fig:DoubleScatCont} includes up to two incoming thin arrows and (always)  two outgoing thin arrows. The incoming and outgoing thin arrows depict the weak (probe) fields received from another atom, and the re-emitted fields, respectively. Since the injected laser field can be strong enough to induce nonlinear inelastic scattering on individual atoms, we furnish thin arrows in Fig.~\ref{fig:DoubleScatCont} with frequency values, which generally differ from the incident laser frequency $\omega_L$. To express the polarization-sensitive character of CBS from vector atoms, we equip the arrows with the polarization indices. In accordance with our selection of the polarization channel [see Sec.~\ref{sec:model}], the laser and the backscattered fields carry indices $+1$ and $-1$, respectively. Then, taking into account that we consider the dipole transition $J_g=0\leftrightarrow J_e=1$, with the non-degenerate ground state, we can unambiguously determine the two polarization indices of the intermediate arrows. Namely, regardless of the arrow's type (solid or dashed), its start and end are supplied with indices $+1$ and $-1$, respectively [see Fig.~\ref{fig:DoubleScatCont}].

Knowledge of the polarization states of the intermediate amplitudes is important when combining single-atom responses into multiple scattering signals. Let a pair of atoms be connected by a solid arrow (positive-frequency field), whose start and end carry the polarization indices $q$ and $q^\prime$, respectively. Then the probability amplitude of the associated double scattering process is proportional to the matrix element \cite{ralf13}
\begin{equation}
\overleftrightarrow{\boldsymbol\Delta}_{q^\prime q}\equiv \hat{\bf e}^*_{q^\prime}\cdot \overleftrightarrow{\boldsymbol\Delta}\cdot \hat{\bf e}_q,
\label{elem_Delta}
\end{equation}
where $\overleftrightarrow{\boldsymbol\Delta}=\overleftrightarrow{\boldsymbol{\mathbbm 1}}-\hat{\textbf{n}} \hat{\textbf n}$ is a
projection operator on the plane transverse to the line connecting the two atoms, with $\overleftrightarrow{\boldsymbol{\mathbbm 1}}$ the identity operator and $\hat{\textbf n}$ the unit vector connecting two atoms. Explicitly,
\begin{eqnarray}
\overleftrightarrow{\boldsymbol{\mathbbm 1}}&=&-\hat{\textbf e}_{-1}\hat{\textbf e}_{+1} +\hat{\textbf e}_{0}\hat{\textbf e}_{0}-\hat{\textbf e}_{+1}\hat{\textbf e}_{-1} \label{eqn:UnityTensor}, \\
\hat{\textbf n}&=&\frac{e^{i \phi} \sin \theta}{\sqrt 2} \hat{\textbf e}_{-1}+\cos \theta \hat{\textbf e}_{0}-\frac{e^{-i \phi} \sin \theta}{\sqrt 2} \hat{\textbf e}_{+1}, \label{eqn:UnitVector}
\end{eqnarray}
with angles $(\theta,\phi)$ which fix the relative orientation of the two atoms (to be averaged over). The probability amplitude of the complex conjugate process (dashed arrow) is proportional to $(\overleftrightarrow{\boldsymbol\Delta}_{q^\prime q})^*=\overleftrightarrow{\boldsymbol\Delta}_{qq^\prime}$. Then, the double scattering process whereupon a pair of atoms is connected by one solid and one dashed arrow with polarization indices $q$ and $q^\prime$, respectively, is proportional to the geometric average
\begin{equation}
 \langle \overleftrightarrow{\boldsymbol\Delta}_{q^\prime q}\overleftrightarrow{\boldsymbol\Delta}_{qq^\prime}\rangle=\frac{1}{4\pi}\int_0^{\pi}d\theta \sin\theta \int_0^{2\pi}d\phi \overleftrightarrow{\boldsymbol\Delta}_{q^\prime q}\overleftrightarrow{\boldsymbol\Delta}_{qq^\prime}. \label{disorder_double}
 \end{equation}

Once the atomic internal structure, the frequencies and the polarizations of all the incoming and outgoing fields are specified, one proceeds with finding the expressions for the building blocks (A)-(D) from perturbative solutions of the generalized OBE (see Sec.~\ref{sec:generalOBE}). As shown in \cite{ralf13}, these blocks can be diagrammatically expanded into the elastic and inelastic spectral response functions. Remarkably, the diagrammatic expansions for vector atoms \cite{ralf13} are the same as those for scalar atoms \cite{SlavaDiagrams}, apart from the arrows' polarization indices, in the vectorial case. Moreover, there is a systematic way of obtaining the analytical expressions for the spectral responses corresponding to scalar \cite{SlavaTriple} and vector \cite{Kneddi} atoms alike (see also Appendix \ref{app:general_expressions}).

Finally, single-atom building blocks are self-consistently reassembled into double scattering diagrams \cite{SlavaDiagrams,ralf13}, from which the mathematical expressions for the frequency-resolved signals are obtained. We will refer to the double scattering ladder and crossed spectra as $L^{(2)}(\nu)$ and $C^{(2)}(\nu)$ ($\nu=\omega_D-\omega_L$), respectively. 

The double scattering contribution to CBS of light from saturated isotropic atoms in the h $\parallel$ h channel has been thoroughly studied before \cite{wellens04,PhysRevLett.94.043603,PhysRevA.73.063813,PhysRevA.76.043832,shatokhin10a,gremaud06,ralf,ralf13}. In particular, the master equation \cite{PhysRevA.76.043832} and the DPP \cite{ralf,ralf13} approaches yield numerically identical results for the double scattering CBS spectra from atoms with isotropic transitions. Therefore, we now move on to the case of triple scattering.

\subsection{Triple scattering}
\label{sec:triple}
\begin{figure}
\begin{center}$
\begin{array}{lr}
\includegraphics[width=3.55in]{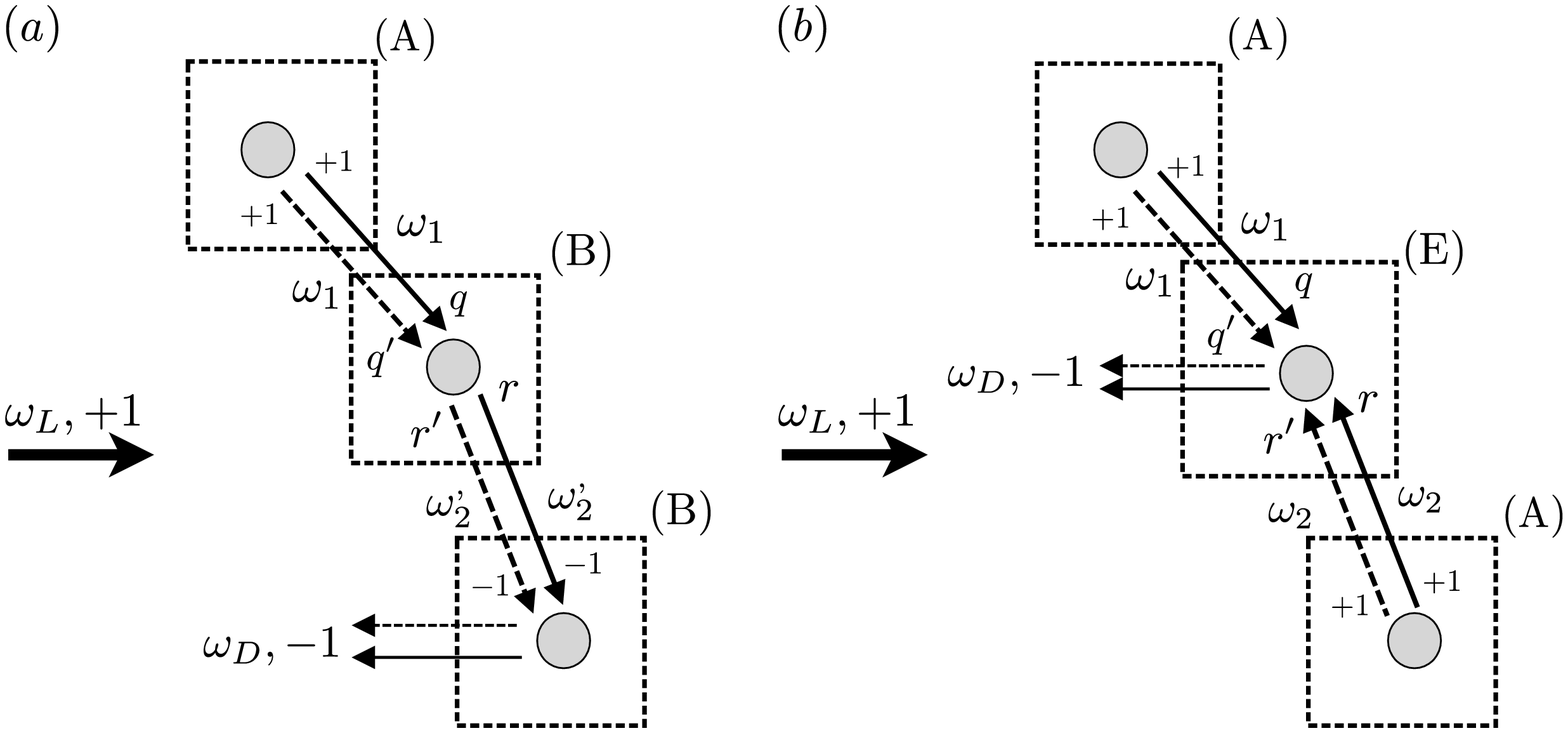}\\
\includegraphics[width=3.55in]{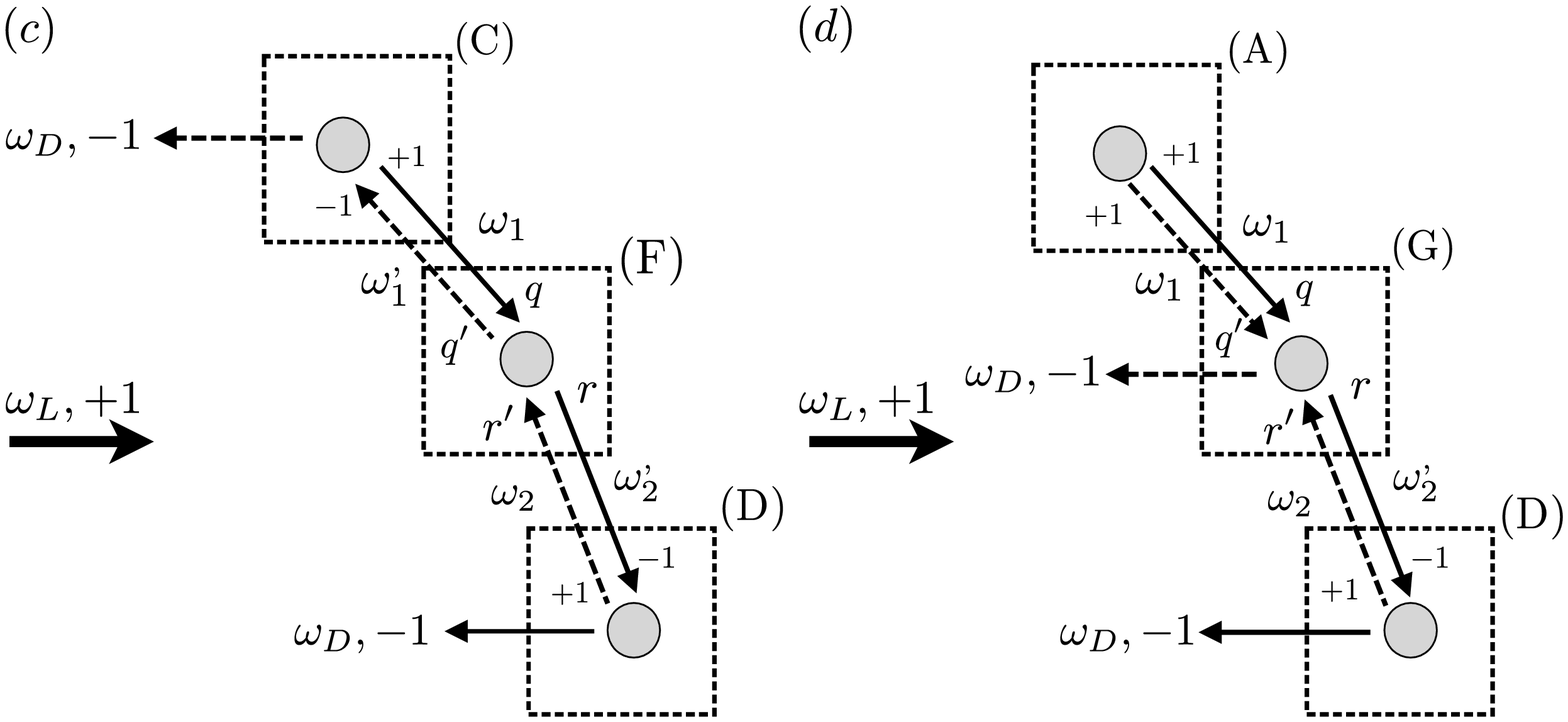}
\end{array}$
\end{center}
\caption{Triple scattering processes surviving the disorder average. The meaning of the notations is the same as in Fig.~\ref{fig:DoubleScatCont}. (a), (b) ladder; (c), (d) crossed contributions.  Each of the diagrams (a)-(d) can be split into three single-atom building blocks (dotted frames). Building blocks (A), (B), (C), (D) and (E), (F), (G) are obtained by solving the generalized OBE under bichromatic and trichromatic driving, respectively.}
\label{fig:TripleScatCont}
\end{figure}

In the triple scattering scenario, all atoms from Fig.~\ref{fig:ThreeSrAtoms} are involved in the scattering sequences. Due to the nonlinearity of the atomic scatterers, there are two types of both, the ladder [see Fig.~\ref{fig:TripleScatCont}(a,b)] and the crossed [see Fig.~\ref{fig:TripleScatCont}(c,d)] diagrams which survive the disorder average. More precisely, if we replace all dashed arrows by solid ones and vice versa in Fig. \ref{fig:TripleScatCont}(d), we obtain another type of the crossed diagram. The latter, however, is the complex conjugate of the one in Fig. \ref{fig:TripleScatCont}(d). Note that Fig.~\ref{fig:TripleScatCont} features the building blocks (A)-(D) that are familiar from the double scattering case (see Fig.~\ref{fig:DoubleScatCont}). In addition, there appear three new building blocks: (E), (F), and (G), which describe spectral responses of the middle atom in Fig.~\ref{fig:TripleScatCont}(a-d). By inspecting the frequencies of the incoming probe fields for these new building blocks, we see that the blocks (E), (F), and (G) require a solution of the OBE for trichromatic driving. 

The diagrams in Fig.~\ref{fig:TripleScatCont}
generalize the already treated case of triple scattering from saturated scalar atoms \cite{SlavaTriple}. As discussed in Sec.~\ref{sec:double}, in the vector-atom case, we equip all the arrows in the diagrams with polarization indices.
However, in contrast to the case of double scattering, selecting the excitation/detection polarization channel does not uniquely specify all the incoming and outgoing fields' polarizations in the triple scattering case. Indeed, polarization indices of the arrows received by the {\it middle} atom in the diagrams in Fig.~\ref{fig:TripleScatCont} can generally take values $\pm 1, 0$.

Some of these combinations do not survive the disorder average and yield vanishing contributions. To identify those combinations, note that the geometric weights for any of the triple scattering diagrams in Fig.~\ref{fig:TripleScatCont}, with fixed polarization indices, are determined analogously as in the case of double scattering, see Eq.~(\ref{disorder_double}). The only difference is that, for triple scattering, the geometric configuration is specified by two independent pairs of random angles. Therefore, the geometric weights for triple scattering factorize into products thereof for double scattering. For example, the geometric weight corresponding to the ladder diagram in Fig.~\ref{fig:TripleScatCont}(a) reads (for brevity, we omit the `+' in the subscripts referring to the polarization index $+1$):
\begin{equation}
\langle\overleftrightarrow{\boldsymbol\Delta}_{q1}\overleftrightarrow{\boldsymbol\Delta}_{1q^\prime}\overleftrightarrow{\boldsymbol\Delta}_{-1r}\overleftrightarrow{\boldsymbol\Delta}_{r^\prime-1}\rangle\!\!=\!\!\langle\overleftrightarrow{\boldsymbol\Delta}_{q1}\overleftrightarrow{\boldsymbol\Delta}_{1q^\prime}\rangle\langle \overleftrightarrow{\boldsymbol\Delta}_{-1r}\overleftrightarrow{\boldsymbol\Delta}_{r^\prime-1}\rangle.
\label{eqn:TripleWeights}
\end{equation}
The geometric weights for the remaining diagrams in Fig.~\ref{disorder_double} are defined analogously.

Using the definitions (\ref{elem_Delta})-(\ref{disorder_double}), it is easy to check that Eq.~(\ref{eqn:TripleWeights}) yields zero, unless $q=q^\prime$ and $r=r^\prime$. From Fig.~\ref{fig:TripleScatCont}(a), we see that these indices describe the polarizations of the co-propagating incoming and outgoing arrows, respectively. Certainly, the same property holds also for the polarization indices which describe the co-propagating amplitudes in Fig.~\ref{fig:TripleScatCont}(b, d). This restriction reduces the number of the building blocks that actually need to be evaluated.

Another restriction on the polarization indices stems not from the configuration average, but rather from basic atom-light interaction considerations.  Namely, for the $J_g=0\leftrightarrow J_e=1$ transition, the polarization of the outgoing field must coincide with the polarization of the incoming laser or of the probe field(s). Otherwise, the spectral responses are identically equal to zero. Taking again as example Fig.~\ref{fig:TripleScatCont}(a), we note that, if $q=q^\prime=+1$, then always $r=r^\prime=+1$, since the laser field is also chosen to have the polarization index $+1$. Accordingly, when $q=q^\prime=0(-1)$, two possibilities emerge: $r=r^\prime=+1$ and $r=r^\prime=0(-1)$.

Once we ensured that our choice of the polarization indices in Fig.~\ref{fig:TripleScatCont} gives rise to a non-zero contribution, we proceed with diagrammatic expansions of the building blocks into elastic and inelastic contributions. This is done in full analogy with the double scattering case (see Sec.~\ref{sec:double}). Using the general formulas given in Appendix \ref{app:general_expressions}, we can then obtain the expressions for the elastic and  inelastic response functions corresponding to the selected polarizations and frequencies of the incoming arrows.

Thereafter, we self-consistently reassemble the individual terms in the diagrammatic expansions of the building blocks into the triple scattering spectral signals, by complete analogy with the case of scalar atoms \cite{SlavaTriple}. Finally, upon summation over all relevant values of the polarization indices, with the corresponding geometric weights [see Eq.~(\ref{eqn:TripleWeights})], we obtain the triple scattering spectra.
We will denote the ladder spectra represented by Figs.~\ref{fig:TripleScatCont}(a) and (b) as $L_1^{(3)}(\nu)$ (type one ladder) and $L_2^{(3)}(\nu)$ (type two ladder), and the crossed spectra represented by Figs.~\ref{fig:TripleScatCont}(c) and (d) as $C_1^{(3)}(\nu)$ (type one crossed) and $C_2^{(3)}(\nu)$ (type two crossed), respectively. 
We note that permutations of three atoms give rise to six type one ladder diagrams, three type two ladde diagrams and six crossed diagrams of each type.  
Then, up to a common prefactor which is absorbed in the statistical weight $w_3$ (see Sec. \ref{sec:total}), the total triple scattering ladder and crossed spectra are given by
\begin{align}
L^{(3)}(\nu)&=L_1^{(3)}(\nu)+\frac{1}{2}L_2^{(3)}(\nu), \label{L3a}\\
C^{(3)}(\nu)&=C_1^{(3)}(\nu)+2 {\rm Re}\{C_2^{(3)}(\nu)\}.
\label{C3a}
\end{align}
Before we present our results in Sec.~\ref{sec:num_results}, we next recall the generalized OBE under classical multi-chromatic driving -- the basic equations used to derive the single atom spectral responses.

\subsection{Generalized optical Bloch equations}
\label{sec:generalOBE}

Let us remind that the single atom spectral responses for double and triple scattering  can be obtained from solutions of the generalized OBE under classical bichromatic (building blocks (A)-(D) in Figs. \ref{fig:DoubleScatCont} and \ref{fig:TripleScatCont}) or trichromatic driving (building blocks (E)-(G) in Fig. \ref{fig:TripleScatCont}). 
Since writing such equations down for an arbitrary number of classical driving fields involves no technical overhead, we here present the generalized OBE for polychromatic driving. As a side remark, we note that the spectral responses for polychromatic classical driving can be used to describe transport of radiation in a dilute medium of saturated atoms \cite{binninger12}.

The generalized OBE can be conveniently obtained from a master equation for the quantum-mechanical expectation value of an arbitrary atomic operator $Q$. We assume that the vector atom is driven by $N+1$ coherent components, of which one represents the laser field with frequency $\omega_L$ and polarization $\hat{\bf e}_L=\hat{\bf e}_{+1}$; and the remaining $N$ components are weak probe fields with frequencies $\omega_1$, $\ldots$, $\omega_N$ and polarizations $\hat{\bf e}_{q_1},\ldots, \hat{\bf e}_{q_N}$, where $q_k=\pm 1,0$ [see Eq.~(\ref{eqn:SphericalBasis})] and $1\leq k\leq N$. In the frame rotating at the laser frequency, the time evolution of the quantum mechanical expectation value $\langle Q\rangle$ is governed by the following master equation \cite{ralf13}:
\begin{align}
\langle\dot{Q}\rangle&\!=\!\left\langle\!\!-i\delta[{\bf D}^\dagger\!\cdot\! {\bf D},\!Q]
\!-\!\frac{i\Omega}{2}[({\bf D}^\dagger\!\cdot\! \hat{\bf e}_{+1})\!+({\bf D}\!\cdot\! \hat{\bf e}_{+1}^*),Q]\right.\nonumber\\
&\left.+\gamma\left({\bf D}^\dagger\!\cdot\![Q,{\bf D}]+[{\bf D}^\dagger,Q]\!\cdot\!{\bf D}\right)\right.\nonumber\\
&\left. -\frac{i}{2}\sum_{k=1}^N[ g_ke^{-i\delta_k t}({\bf D}^\dagger\!\cdot\! \hat{\bf
e}_{q_k})
+ g_k^*e^{i\delta_kt}({\bf D}\!\cdot\! \hat{\bf
e}^*_{q_k}),Q]\right\rangle. \label{eq:master_vector}
\end{align}
Here, $\delta=\omega_L-\omega_0$ and $\Omega=2d{\cal E}_L/\hbar$ is the  (real) Rabi frequency, which describes the coupling of the laser field to the transition with a magnetic quantum number $m=1$. The, $g_k$ are the weak ($|g_k|\ll \gamma$) probe-field Rabi frequencies, and $\delta_k=\omega_k-\omega_L$ the probe-laser field detunings. Finally, ${\bf D}$ and ${\bf D}^\dagger$ are the atomic lowering and raising vector operators, respectively. They define the atomic dipole operator through ${\boldsymbol{\cal D}}=d({\bf D}^\dagger+{\bf D})$. In the spherical basis, the dipole lowering operator corresponding to the $J_g=0\leftrightarrow J_e=1$ transition can be expanded as
\begin{equation}
{\bf D}=-\hat{\bf e}_{-1}\sigma_{12}+\hat{\bf e}_0\sigma_{13}-\hat{\bf e}_{+1}\sigma_{14},
\label{dplus_general}
\end{equation}
where $\sigma_{ij}\equiv |i\rangle\langle j|$, and $|1\rangle$ ($|3\rangle$) is the ground (excited) state (sub)level with magnetic quantum number $m=0$, whereas $|2\rangle$ and $|4\rangle$ are the excited state sublevels  with magnetic quantum numbers $m=-1$ and $m=1$, respectively.

By choosing operators $Q$ from the complete orthonormal set of operators for the isotropic transition (see, for instance, \cite{PhysRevA.76.043832}), we translate Eq. (\ref{eq:master_vector}) into the generalized OBE under polychromatic classical driving:
\begin{align}
\langle\dot{\textbf{Q}}(t)\rangle&={\bf M}\langle\textbf Q(t)\rangle+{\bf L}
+\sum_{k=1}^N[e^{-i\delta_k t}\boldsymbol{\Delta}_{q_k}^{(-)}\langle{\bf Q}(t)\rangle
\nonumber \\&
+e^{i\delta_k t}\boldsymbol{\Delta}_{q_k}^{(+)}\langle{\bf Q}(t)\rangle],\label{Q_n}
\end{align}
where the matrix ${\bf M}$ describes the dipole's radiative decay, as well as its coupling to the laser field; the matrix $\boldsymbol{\Delta}_{q_k}^{(-)}$  ($\boldsymbol{\Delta}_{q_k}^{(+)}$), proportional to $g_k$ ($g_k^*$), describes the coupling of the atom to the positive- (negative-)frequency probe-field component with polarization $q_k$. The explicit form of the $15\times 15$ matrices ${\bf M}$ and $\boldsymbol{\Delta}_{q_k}^{(\pm)}$, together with the 15-dimensional vector ${\bf L}$, can readily be defined using Eq.~(\ref{eq:master_vector}), once the vector $\langle{\bf Q}\rangle$ is specified.

General perturbative analytical solutions of Eq. (\ref{Q_n}) are given in Appendix \ref{app:general_expressions}. In the next section, we present our results that are obtained after a self-consistent combination of single atom responses, deduced from these solutions, into the triple scattering signals. 

\begin{figure}[t!]
\includegraphics[width=0.48\textwidth]{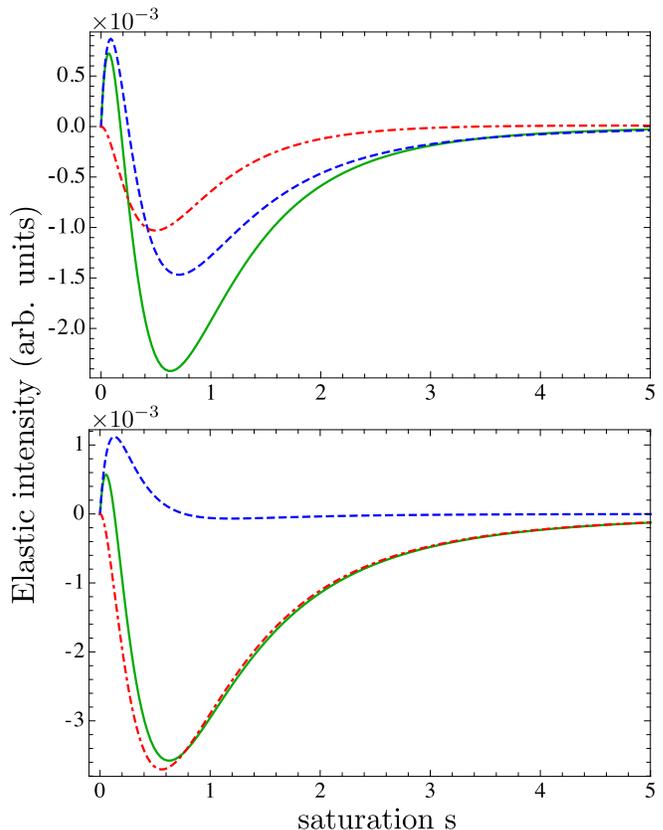}
\caption{(color online) Triple scattering elastic ladder (top) and
crossed (bottom) intensities, together with their decompositions
into type one and type two components, see Eqs.~(\ref{exp_L}),
(\ref{exp_C}), at resonant driving as a function of saturation $s$.
(Solid lines) total elastic signals, $L^{(3)}_{el}$ and
$C^{(3)}_{el}$; (dashed lines) type one ladder and crossed elastic
components, $L^{(3)}_{el,1}$ and $C^{(3)}_{el,1}$, corresponding to diagrams in
Fig.~\ref{fig:TripleScatCont}(a) and (c), respectively; (dashed-dotted
lines) type two ladder and crossed elastic components, $L^{(3)}_{el, 2}/2$ and $2{\rm Re}\{C^{(3)}_{el, 2}\}$, corresponding
to diagrams in Fig.~\ref{fig:TripleScatCont}(b) and (d),
respectively.} \label{fig:ElInt}
\end{figure}

\begin{figure*}[!t]
\includegraphics[width=\textwidth]{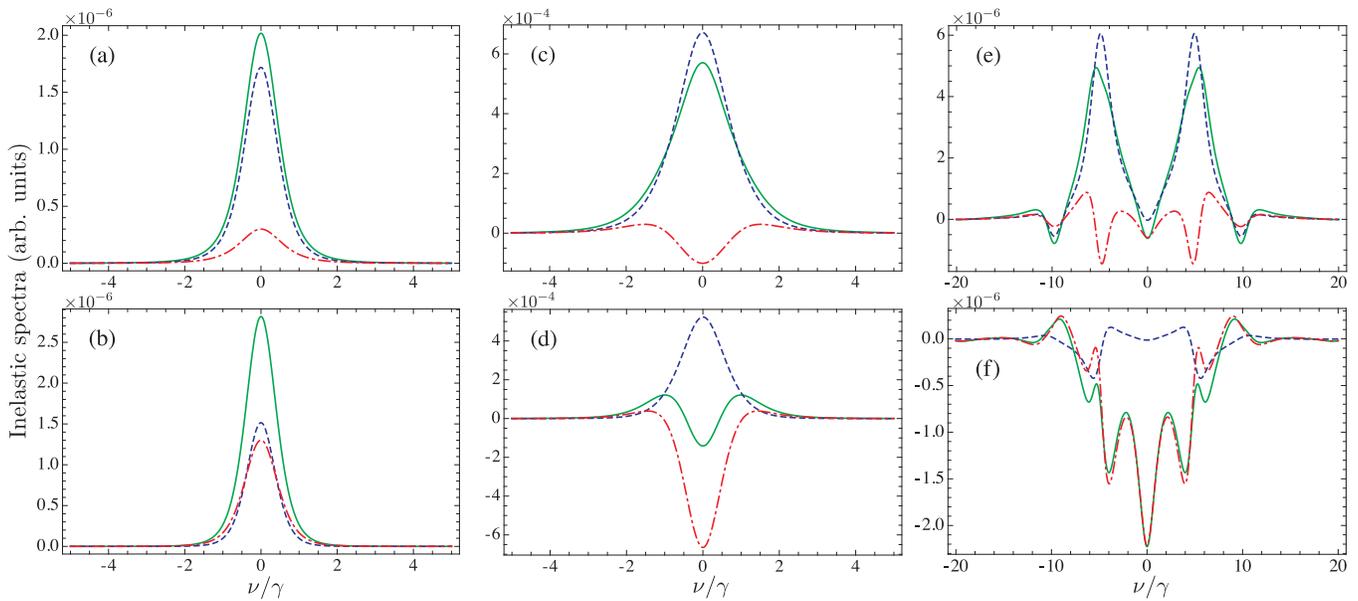}
\caption{(color online) Examples of the triple scattering inelastic ladder (top) and
crossed (bottom) spectra, together with their decompositions
into type one and type two components, see Eqs.~(\ref{exp_L}),
(\ref{exp_C}), at resonant driving and for different values of the laser Rabi frequency:
 (a,b) $\Omega=0.1 \gamma$, (c,d) $\Omega=1.0\gamma$, (e,f) $\Omega=10.0 \gamma$. (Solid lines) total inelastic spectra, $L^{(3)}_{in}(\nu)$ and
$C^{(3)}_{in}(\nu)$; (dashed lines) type one ladder and crossed inelastic
components, $L^{(3)}_{in,1}(\nu)$ and $C^{(3)}_{in,1}(\nu)$, corresponding to diagrams in Fig.~\ref{fig:TripleScatCont}(a) and (c), respectively; (dashed-dotted
lines) type two ladder and crossed inelastic components, $L^{(3)}_{in, 2}(\nu)/2$ and $2{\rm Re}\{C^{(3)}_{in, 2}(\nu)\}$, corresponding
to diagrams in Fig.~\ref{fig:TripleScatCont}(b) and (d),
respectively.}
\label{fig:InelSpec}
\end{figure*}

\section{Numerical results}
\label{sec:num_results}
  \subsection{Triple scattering spectra} \label{sec:TripleSpectra}

Let us consider the triple scattering spectra, $L^{(3)}(\nu)$ and $C^{(3)}(\nu)$ [see Eqs.~(\ref{L3a}), (\ref{C3a})], in different regimes of the laser-atom interaction. In general, the spectra can be decomposed into {\it el}astic and {\it in}elastic components,
\begin{align}
L^{(3)}(\nu)&=L^{(3)}_{el}(\nu)+L^{(3)}_{in}(\nu),\label{L3aa}\\
C^{(3)}(\nu)&=C^{(3)}_{el}(\nu)+C^{(3)}_{in}(\nu).\label{C3aa}
\end{align}
Furthermore, each term on the right hand sides of
Eqs.~(\ref{L3aa}), (\ref{C3aa}) can be expanded into a sum of the
type one and two components (see Eqs. (\ref{L3a}), (\ref{C3a})):
\begin{align}
L^{(3)}_{el(in)}(\nu)&=L^{(3)}_{el(in),1}(\nu)+\frac{1}{2}L^{(3)}_{el(in),2}(\nu),\label{exp_L}\\
C^{(3)}_{el(in)}(\nu)&=C^{(3)}_{el(in),1}(\nu)+2 {\rm Re}\{C^{(3)}_{el(in),2}(\nu)\}.\label{exp_C}
\end{align}

Here, we study the behavior of the elastic and inelastic spectra
for different laser-field Rabi frequencies $\Omega$. Since the
atomic response to the external driving depends also on the detuning
$\delta$ between the laser and the atomic transition frequencies, it
is convenient to use the saturation parameter,
\begin{eqnarray}
s=\frac{1}{2} \frac{\Omega^2}{\delta^2+\gamma^2},
\label{eqn:SatPara}
\end{eqnarray}
to describe different regimes of atom-laser interaction \cite{cohen1998atom}. In particular,
the elastic and inelastic scattering regimes are characterized by the inequalities $s\ll1$ and $s\geq1$,
respectively. We will see below that, in the inelastic regime, the triple scattering spectra
can exhibit negative values. This is not unphysical, because triple scattering is only one
contribution (among double and higher order contributions) to the total multiple scattering signal,
which is strictly positive.

  \subsubsection{Elastic spectrum}
When we speak of the elastic spectra, we refer to the monochromatic components that are scattered at
the laser frequency. The elastic triple scattering ladder and crossed spectra are then proportional
to delta-functions: $L^{(3)}_{el}\delta(\nu)$ and $C^{(3)}_{el}\delta(\nu)$, where the
components' intensities, $L^{(3)}_{el}$ and $C^{(3)}_{el}$, depend on the saturation parameter.

Figure~\ref{fig:ElInt} shows a plot of the elastic triple scattering ladder and crossed
intensities as a function of $s$. The oscillatory behavior of the ladder and crossed components
is due to the opposite contributions of the individual ladder and crossed type one and two processes,
respectively. For $s\ll1$, the elastic ladder and crossed intensities are positive and grow linearly with $s$ (see solid lines in Fig. \ref{fig:ElInt}). 
Further on, the ladder and crossed elastic intensities, are increasing with $s$ until $s\approx 0.1$. For larger values of $s$, nonlinear scattering
leads to a decrease of the elastic intensities, which yields negative contributions for
$s\geq 0.2$. In the deep saturation regime, $s\gg1$, as expected, the elastic intensities
tend to zero (remaining negative).

  \subsubsection{Inelastic spectrum}

In contrast to the elastic components, the inelastic triple scattering ladder and
crossed spectra, $L^{(3)}_{in}(\nu)$ and $C^{(3)}_{in}(\nu)$, are emitted over a range of
frequencies. In Fig.~\ref{fig:InelSpec}, we present several examples of inelastic spectra at
$\delta=0$, and for different values of the laser Rabi frequency $\Omega$.

In the weakly inelastic scattering regime, the inelastic spectra are dominated by two-photon processes \cite{SlavaTriple}. At exactly resonant driving, the ladder and crossed spectra then consist of a single peak centered at the laser frequency, i.e., at $\nu=0$ [see Fig.~\ref{fig:InelSpec}(a) and (b)]. Furthermore, the crossed component of the inelastic spectrum has a larger maximum than the ladder one in this limit. As noted in \cite{PhysRevA.73.013802}, this can lead to a CBS enhancement factor larger than two, provided that the elastic component is filtered out.

At larger values of $\Omega$, the increasing influence of higher order inelastic multi-photon processes leads to a reduction of the enhanced backscattering. Partially, this happens owing to the opposite interference character of  type one and type two crossed components [see Fig.~\ref{fig:InelSpec}(d) and (f)]. Moreover, in this limit, the spectra split into several Lorentzian and dispersive resonances [see Fig.~\ref{fig:InelSpec}(e) and (f)]. The number and positions thereof can be understood from the dressed-state structure of the relevant dipole transitions of the atoms, which are involved in the corresponding triple scattering processes, in full analogy to the double scattering case \cite{PhysRevA.76.043832}.

\subsection{Total CBS signal}\label{sec:total}
\begin{figure}[b!]
\includegraphics[width=0.48\textwidth]{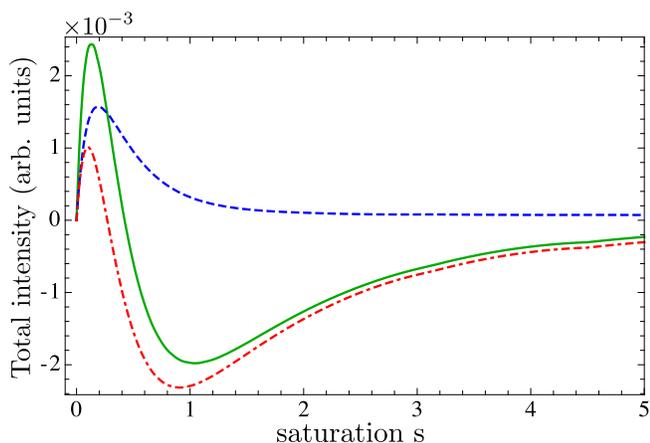}
\caption{(color online) Total triple scattering ladder (dashed) and
crossed (dashed-dotted) intensities, $L^{(3)}_{tot}$ and
$C^{(3)}_{tot}$, respectively, see Eq. (\ref{C3}),
together with the total triple scattering CBS signal, $L^{(3)}_{tot}+C^{(3)}_{tot}$
(solid line), at resonant driving, as a function of the saturation
$s$.} \label{fig:TotTripInt}
\end{figure}
Integration of the double and triple scattering spectra (see Sec.~\ref{sec:TripleSpectra}) over their frequency distributions yields the total ladder and crossed intensities of the corresponding order:
\begin{equation}
L_{tot}^{(j)}=\int_{-\infty}^{\infty}d\nu L^{(j)}(\nu),\quad 
C_{tot}^{(j)}=\int_{-\infty}^{\infty}d\nu C^{(j)}(\nu),\label{C3} 
\end{equation}
where $j=2,3$. The total double scattering intensities
have been discussed in detail in Refs.~\cite{PhysRevLett.94.043603,PhysRevA.73.063813}. We therefore move on to the triple scattering case.

Our plots of $L^{(3)}_{tot}$ and $C^{(3)}_{tot}$ as functions of the saturation parameter $s$ are presented in
Fig.~\ref{fig:TotTripInt}. In the elastic scattering regime ($s\ll1$),
the corresponding ladder and crossed components show a monotonic
increase, and reach a maximum around $s\approx0.15$, when inelastic
processes start to set in. Further increase of $s$ results in a
monotonic decrease of the ladder component as $s^{-2}$. The crossed component features
destructive interference character in the saturation regime, with a
minimum at $s\approx 0.9$. As the ladder intensity, the crossed components decay quadratically to zero, for large $s$.

With the total double and triple scattering intensities being defined, we calculate the total ladder and crossed intensities in an optically thin cloud of cold atoms as
\begin{subequations}
\begin{align}
L_{tot}&=w_2 L_{tot}^{(2)}+w_3 L_{tot}^{(3)}, \label{eqn:IntLad} \\
C_{tot}&=w_2 C_{tot}^{(2)}+w_3C_{tot}^{(3)} \label{eqn:IntCros},
\end{align}
\label{CL_tot}
\end{subequations}
where $w_2$ and $w_3$ are the statistical weights which determine the
fractions of the double and triple scattering contributions,
respectively. Since the experimental values $w_2$ and $w_3$ are
unavailable, we obtain them on the basis of a Monte Carlo
simulation procedure, as described in the next section.

\label{sec:results}
  \subsection{Determination of the weights $w_2$ and $w_3$}
  \label{sec:DTweights}
To estimate the statistical weights $w_2$ and $w_3$ [see Eq.~(\ref{CL_tot})], we employ a Monte Carlo method. We simulate a random walk of a ``particle'' inside a sphere confining randomly distributed point ``scatterers''. This roughly corresponds to multiple elastic scattering of a photon in an atomic cloud.
In accordance with the experimental parameters \cite{PhysRevE.70.036602}, we choose the radius of the sphere and the scattering mean free path to be 0.7 mm and 0.75 mm, respectively. As noted in \cite{PhysRevE.70.036602}, these values are in agreement with the measured optical thickness of 3.5.

A simulation of the random walk consists in generating locations of the scattering events inside the sphere, with an exponential step length distribution. The latter ensures consistency of the walk with the radiation transfer equation \cite{Heiderich1994110}.
After repeating this procedure
one million times, the percentage of single, double, etc. scattering events converges to their
constant values $w_1$, $w_2$, etc., respectively.

We note that this procedure ignores the scattering properties of the atomic dipoles. These properties may, for instance, lead to different dependences of the statistical weights on $s$ in the saturation regime. However, we expect that for not too strong laser fields (as in the experiment \cite{PhysRevE.70.036602}), the CBS signal is still dominated by the elastic component. In this case, the relative weights of different scattering orders should approximately be equal to those in a realistic atomic cloud having the same root mean square radius and scattering mean free path. 

We found that the statistical weights of single, double, triple, and multiple ($>3$) scattering orders are $56\%$,
$23\%$, $11\%$, and  $10\%$, respectively. As expected, single scattering dominates owing to the small optical thickness of the cloud. However, this contribution is filtered out in the helicity preserving channel and does not enter Eq.~(\ref{CL_tot}). Furthermore, the inequality $w_2>w_3$ which holds for optically thin clouds, together with the positivity of the total double scattering intensity \cite{PhysRevLett.94.043603}, ensures positivity of the total signal including double and triple scattering contributions. In particular, in the linear scattering regime ($s\ll1$), where the inelastic intensity scales as $\sim s^2$ and thus can be neglected, we recover the positivity of each scattering order (double and triple scattering) by itself, as evident from the behavior of the elastic intensity in Fig.~\ref{fig:ElInt}. Positivity of the signal, in turn, justifies dropping higher scattering orders ($>3$), whose evaluation requires more than three atoms, and is beyond the scope of this work. As a concluding remark, we note that filtering out the single scattering contribution leads to a renormalization of the weights $w_2$ and $w_3$, but does not change their ratio. Therefore, in Eq.~(\ref{CL_tot}), we substitute $w_2=0.23$ and $w_3=0.11$. 

\begin{figure}[b!]
\includegraphics[width=0.48\textwidth]{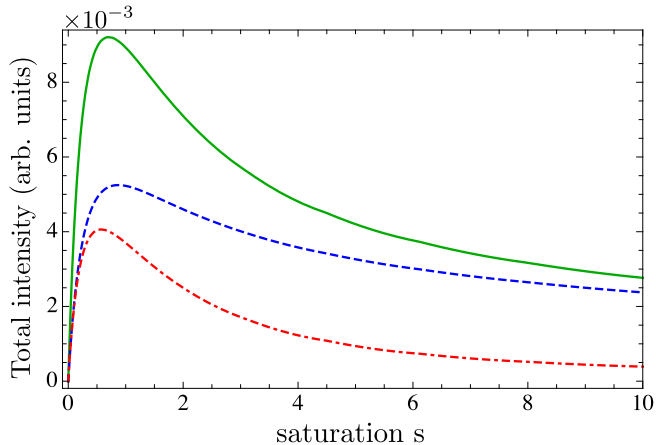}
\caption{(color online) Total ladder (dashed) and crossed (dashed-dotted) intensities, $L_{tot}$ and $C_{tot}$, respectively, see Eq. (\ref{eqn:IntLad}) and (\ref{eqn:IntCros}), and the corresponding total CBS signal, $L_{tot}+C_{tot}$ (solid line), for a mixture of double scattering contributions, taken with the weights $0.23$ and $0.11$, respectively, at resonant driving, as a function of the saturation parameter $s$.} \label{fig:TotInt}
\end{figure}

\subsection{Enhancement factor}\label{sec:enh}
This section is devoted to the presentation and discussion of our numerical results for the CBS enhancement factor as a function of the saturation parameter $s$. Given the total ladder and crossed intensities, Eq.~(\ref{CL_tot}), the enhancement factor is defined as
\begin{equation}
\alpha=1+\frac{C_{tot}}{L_{tot}}.
\label{eqn:EnhFactor}
\end{equation}
Thus, the results for the total double and triple scattering intensities,
presented in Fig.~\ref{fig:TotInt}, allow us to directly determine the corresponding
behavior of the enhancement factor $\alpha$.

\begin{figure}[t!]
\includegraphics[width=0.47\textwidth]{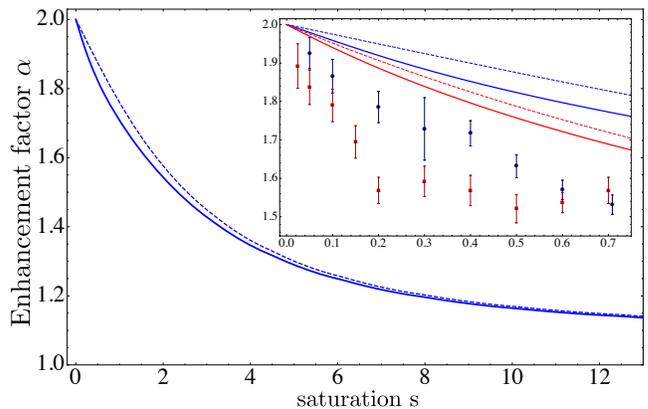}
\caption{(color online) Enhancement factor $\alpha$ vs. saturation $s$ at resonant driving for the double scattering contribution  (dashed line),  and for a mixture of the double and triple scattering contributions, taken with the weights $0.23$ and $0.11$, respectively (solid line). Inset: comparison of the calculated and experimental enhancement factor (dots with error bars, from Ref. \cite{PhysRevE.70.036602}),  in the range $0\leq s\leq 0.8$, for detunings $\delta=0$ (blue) and $\delta=\gamma$ (red).}
\label{fig:EnhDoubleDT0k0}
\end{figure}

In Fig. \ref{fig:EnhDoubleDT0k0}, we present a plot of $\alpha$ vs.~$s$ in
the case of exact resonance ($\delta=0$). Dashed lines in Fig.~\ref{fig:EnhDoubleDT0k0}
represent the enhancement factor for the double scattering contribution, first obtained in
 \cite{PhysRevLett.94.043603}, and solid lines show our present results for the double and
 triple scattering contributions derived using Eqs.~(\ref{CL_tot})-(\ref{eqn:EnhFactor}).
 We see that an account of the double and triple scattering contributions leads to a faster
 decay of the enhancement factor with increasing saturation of the atoms, than by considering
 double scattering alone. However,  this speedup of coherence loss which is noticeable at
 small $s$, becomes negligible in the saturation regime. As seen from Fig.~\ref{fig:EnhDoubleDT0k0},
 our results for the combined double and triple scattering signal, and that of the purely double scattering one, almost merge for $s\geq 8$. This happens because, in the saturation regime, the triple scattering intensity decays faster (as $s^{-2}$, see Sec. \ref{sec:total}) than the double scattering intensity (which decays as $s^{-1}$ \cite{PhysRevLett.94.043603}). 

Even for small $s$, the initial slope of $\alpha(s)$, due to double and triple scattering, is not steep enough to provide a good quantitative 
agreement with the experimental observations
\cite{PhysRevE.70.036602}. To show this, in
Fig.~\ref{fig:EnhDoubleDT0k0} we provide the inset with a magnified
plot of $\alpha$ vs. $s$ in the range $0\leq s \leq 0.8$. This range
covers the transition from the linear scattering regime to the
saturation regime studied in \cite{PhysRevE.70.036602}. In this
experiment, the enhancement factor was measured at resonant driving
and for small laser detuning ($\delta=\gamma$). Numerical results
for the latter are depicted in Fig.~\ref{fig:EnhDoubleDT0k0} by red
lines. We see that, in both cases, of the resonant and detuned
driving, the account of the double and triple scattering
contributions leads to a faster decay of the enhancement factor with
increasing saturation parameter, and to a better agreement with the
experiment than by considering double scattering alone. Despite that, there remain significant deviations between the
experimentally observed and the here predicted behaviors of the enhancement factor.

One possible reason for this is the missing account for multiple scattering processes of higher-than-third order which, according to our simulations of the random walk, may add contributions that are comparable in magnitude to triple scattering. Another reason may be the role of the atomic medium providing a mean-free path, which itself depends on frequency and saturation. It remains to be seen in future work whether a generalization of our approach to treat high scattering orders in the effective medium can lead to a better quantitative agreement between theoretical and experimental results. 

\section{Conlusion} \label{sec:conclusion}
We have studied coherent backscattering of intense laser light from three atoms
with a $J_g=0\leftrightarrow J_e=1$ transition, using the diagrammatic
pump-probe approach. We have been motivated by the need for an improved quantitative description of relevant experiments \cite{PhysRevE.70.036602},
as well as by the fundamental interest in the role of the higher scattering orders in
the inelastic scattering regime.

By combining self-consistently single-atom spectral responses to a classical mono-, bi- and trichromatic fields, we identified those double and triple scattering processes which
survive the disorder average and contribute to the average backscattered light intensity.
The expressions for the corresponding single-atom responses were derived analytically for the general case of $n$ weak probe field components, and subsequently applied to the particular situation described above.
We presented numerical results for the triple scattering elastic and inelastic spectra, and for the total intensity, respectively. Furthermore, to obtain the total detected double and triple scattering signal that is emitted
from a dilute cloud of cold atoms, we deduced the statistical weights of the double and
triple scattering contributions using a classical Monte Carlo simulation of a photon random
walk inside a sphere containing point scatterers, such that the simulated medium's optical thickness
corresponded to the experimental one reported in \cite{PhysRevE.70.036602}.

One of the main quantities of interest in this work was the CBS enhancement
factor as a function of the saturation parameter. We showed that the
enhancement factor deduced from the double and triple scattering
signals exhibits a faster decay as a function of the saturation
parameter, and yields a better qualitative agreement with the
experimental observation \cite{PhysRevE.70.036602} than the
enhancement factor based on the double scattering alone. Yet, the
experimentally observed enhancement factor still decays considerably faster than the
one calculated in this work.

The remaining mismatch between the experimental observation and
theoretical prediction suggests a potential direction of future
research: It would be both important and challenging to
explore the truly multiple scattering regime in a cold atomic gas of
saturated atoms -- a goal which is within reach in the
framework of the diagrammatic pump-probe approach \cite{binninger12}.  \\

\acknowledgements
Discussions with Felix Eckert are gratefully acknowledged.
This work was financially supported by DFG through grant BU-1337/9-1.\\

\appendix

\section{General expressions for single-atom spectral responses}\label{app:general_expressions}
\subsection{Elastic spectral responses} \label{sec:elastic}
\subsubsection{General solution}
\label{app:gen_sol}
The elastic spectral responses are obtained directly from the perturbative solutions to Eq. (\ref{Q_n}). As shown in \cite{Kneddi}, the $n$th-order ($0< n\leq 2N$) correction to the generalized Bloch vector, $\expec{\textbf Q(\delta_1^{[q_1]},\ldots,\delta_n^{[q_n]})}^{(s_1\ldots s_n)}$ ($s_k=+/-$ corresponds to the negative-/positive-frequency character of the $k$th probe field), reads
\begin{align}
    \expec{&\textbf Q(\delta_1^{[q_1]},\ldots,\delta_n^{[q_n]})}^{(s_1\ldots s_n)} \nonumber \\
    &\!=\!\!\!\! \sum\limits_{\pi(j_1,\ldots,j_n)} \textbf G\left(i \zeta\right) \boldsymbol{\Delta}^{(s_{j_n})}_{q_{j_n}}\ldots \textbf G(is_{j_1} \delta_{j_1} \nonumber \\
    &+i s_{j_2} \delta_{j_2}) \boldsymbol{\Delta}^{(s_{j_2})}_{q_{j_2}} \textbf G(i s_{j_1} \delta_{j_1}) \boldsymbol{\Delta}^{(s_{j_1})}_{q_{j_1}} \langle {\bf Q}\rangle^{(0)}. \label{eqn:NthSolOBE}
    \end{align}
Here, $\textbf G(z)=1/(z-\textbf M)$ is the Green's matrix governing the internal dynamics of the laser-driven atom [see Eq. (\ref{Q_n})], $z\equiv z^\prime+iz^{\prime\prime}$ is the Laplace transform variable, $\expec{\textbf Q}^{(0)}=\textbf G(0) \textbf L$ is the zeroth-order solution to Eq. (\ref{Q_n}), $\pi(j_1,\ldots,j_n)$ denotes $n!$ permutations of the indices $j_1,\ldots,j_n\in \{1,\ldots, n\}$, and
\begin{equation}
\zeta\equiv\sum_{k=1}^n{s_{k}\delta_{k}}.
\label{def_lambda}
\end{equation}
The elastic spectral responses are the outgoing positive- and negative-frequency amplitudes represented by two elements of the vector $\expec{\textbf Q(\delta_1^{[q_1]},\ldots,\delta_n^{[q_n]})}^{(s_1\ldots s_n)}$. These elements are selected through scalar products of the vector  $\expec{\textbf Q(\delta_1^{[q_1]},\ldots,\delta_n^{[q_n]})}^{(s_1\ldots s_n)}$ with the corresponding projection vectors \cite{ralf13}. In our derivations of the triple scattering spectra, we have used the solution (\ref{eqn:NthSolOBE}) for up to four incoming probe fields.

Below, we present the proof of Eq. (\ref{eqn:NthSolOBE}).

\subsubsection{Proof of Eq. (\ref{eqn:NthSolOBE})}
\label{app:ProofOBV}
To this end, we use the method of induction.
For $n=1$, that is, for one probe-field component specified by a set $(\delta_{j_1}^{[q_{j_1}]}, s_{j_1})$ (see Sec.~\ref{sec:elastic}), where $j_1\in \{1,\ldots, n\}$, Eq.~(\ref{eqn:NthSolOBE}) reduces to 
\begin{equation}
{\expec{{\textbf Q}(\delta_{j_1}^{[q_{j_1}]})}^{(s_{j_1})}}=\textbf G(i s_{j_1} \delta_{j_1}) \boldsymbol\Delta^{(s_{j_1})}_{q_{j_1}} \expec{\textbf Q}^{(0)},
\end{equation}
which coincides with the solution obtained in \cite{ralf13}.

Next, we need to show that Eq.~(\ref{eqn:NthSolOBE}) holds provided that it holds for $(n-1)$ probe-field components:
 \begin{align}
	\expec{\textbf Q(&\delta_{j_1}^{[q_{j_1}]},\ldots,\delta_{j_{n-1}}^{[q_{j_{n-1}}]})}^{(s_{j_1}\ldots s_{j_{n-1}})}  \nonumber \\
	&\!=\!\!\!\! \sum\limits_{\pi(j_1,\ldots,j_{n-1})} \textbf G\left(i \zeta_n\right) \boldsymbol{\Delta}^{(s_{j_{n-1}})}_{q_{j_{n-1}}}\ldots \textbf G(is_{j_1} \delta_{j_1} \nonumber \\
	&+i s_{j_2} \delta_{j_2}) \boldsymbol{\Delta}^{(s_{j_2})}_{q_{j_2}} \textbf G(i s_{j_1} \delta_{j_1}) \boldsymbol{\Delta}^{(s_{j_1})}_{q_{j_1}} \langle {\bf Q}\rangle^{(0)}, \label{eqn:N-1thSolOBE} 
	\end{align}
where $\zeta_n\equiv \zeta-s_{j_n}\delta_{j_n}$ and $j_1,\ldots,j_{n-1}\in \{1,\ldots, n\}$.

To this end, we seek perturbative solutions of the generalized OBE describing the dynamics of a single atom subjected to $n$ weak probe field components [compare to Eq. (\ref{Q_n})]:
\begin{align}
	\expec{\dot{\textbf Q}(t)}=\textbf M \expec{\textbf Q(t)}+\textbf L + \sum_{i=1}^{n} e^{is_i \delta_{i}t}\boldsymbol \Delta^{(s_i)}_{q_i}\expec{\textbf Q(t)}.
	 \label{eqn:PolyOBE} 
	\end{align}
Analogously to the case of double scattering \cite{Geiger2010244}, the perturbative solution to Eq.~(\ref{eqn:PolyOBE}) is given by the following expansion:
\begin{widetext}
 \begin{eqnarray}
	\expec{\textbf Q(t)}&=&\expec{\textbf Q(t)}^{(0)}+\sum_{i=1}^n e^{ i s_{i}\delta_{i}t} \expec{\textbf Q(\delta_i^{[q_i]};t)}^{(s_{i})} \nonumber \\
	&+&\sum_{\substack{\pi(j_1,j_{2}),\\ j_1< j_{2}}}  e^{i(s_{j_1}\delta_{j_{1}}+s_{j_{2}}\delta_{j_{2}})t} \expec{\textbf Q(\delta_{j_1}^{[q_{j_1}]},\delta_{j_{2}}^{[q_{j_{2}}]};t)}^{(s_{j_1}s_{j_2})}  \nonumber \\
	&+&\ldots \nonumber \\
	&+&\sum_{\substack{\pi(j_1,...,j_{n-1}),\\ j_1<....< j_{n-1}}} e^{i(s_{j_1}\delta_{j_{1}}+\ldots+s_{j_{n-1}}\delta_{j_{n-1}})t}  \expec{\textbf Q(\delta_{j_1}^{[q_{j_1}]},\ldots,\delta_{j_{n-1}}^{[q_{j_{n-1}}]};t)}^{(s_{j_1}\ldots s_{j_{n-1}})} \nonumber \\
	&+& e^{i\zeta t} \expec{\textbf Q(\delta_1^{[q_1]},\ldots,\delta_{n}^{[q_{n}]};t)}^{(s_1\ldots s_{n})}. \nonumber \\
	 \label{eqn:genBlochExpansionN} 
	\end{eqnarray}
\end{widetext}
where $j_1,\ldots,j_{n-1}\in \{1,\ldots, n\}$, and the ordering of indices in the   above sums is to avoid repetitions (recall that $\expec{\textbf Q(\delta_{j_1}^{[q_{j_1}]},\ldots,\delta_{j_{n-1}}^{[q_{j_{n-1}}]};t)}^{(s_{j_1}\ldots s_{j_{n-1}})}$ are fully symmetric with respect to permutations of all indices, see Eq. (\ref{eqn:N-1thSolOBE})) . The subsequent terms on the right hand side of Eq.~(\ref{eqn:genBlochExpansionN}) represent zeroth-, first-, ..., $n$th-order time-dependent solutions of Eq.~(\ref{eqn:PolyOBE}). We focus on the $n$th-order term $\expec{\textbf Q(\delta_1^{[q_1]},\ldots,\delta_{n}^{[q_{n}]};t)}^{(s_1\ldots s_{n})}$, whose steady-state form (\ref{eqn:NthSolOBE}) we need to prove. Inserting Eq.~(\ref{eqn:genBlochExpansionN}) into (\ref{eqn:PolyOBE}) yields the following equation of motion for $\expec{\textbf Q(\delta_1^{[q_1]},\ldots,\delta_{n}^{[q_{n}]};t)}^{(s_1\ldots s_{n})}$:
 \begin{align}
	&\expec{\dot{\textbf{Q}}(\delta_1^{[q_1]},\ldots,\delta_n^{[q_n]};t)}^{(s_1\ldots s_n)} \nonumber \\
	 &= \left(-i\zeta +\textbf M\right) \expec{\textbf Q(\delta_1^{[q_1]},\ldots,\delta_n^{[q_n]};t)}^{(s_1\ldots s_n)} \nonumber \\
	&+ \boldsymbol \Delta^{(s_1)}_{q_1}  \expec{\textbf Q(\delta_2^{[q_2]},\ldots,\delta_n^{[q_n]};t)}^{(s_2\ldots s_n)} \nonumber \\
	&+ \boldsymbol \Delta^{(s_2)}_{q_2}  \expec{\textbf Q(\delta_1^{[q_1]},\delta_3^{[q_3]},\ldots,\delta_n^{[q_n]};t)}^{(s_1s_3\ldots s_n)} \nonumber \\
	&+\ldots \nonumber \\
	&+ \boldsymbol \Delta^{(s_n)}_{q_n}  \expec{\textbf Q(\delta_1^{[q_1]},\ldots,\delta_{n-1}^{[q_{n-1}]};t)}^{(s_1\ldots s_{n-1})}
	 \label{eqn:NthorderOBE} .
	\end{align}
In the steady-state limit, the left hand side of Eq.~(\ref{eqn:NthorderOBE}) vanishes, and we arrive at the following solution for the $n$th-order correction:
\begin{align}
	\expec{\textbf Q(&\delta_1^{[q_1]},\ldots,\delta_n^{[q_n]})}^{(s_1\ldots s_n)} \nonumber \\
	=& \textbf G\left(i\zeta\right) \left[  {\boldsymbol\Delta}^{(s_{1})}_{q_{1}}\expec{\textbf Q(\delta_2^{[q_2]},\ldots,\delta_n^{[q_n]})}^{(s_2\ldots s_n)} \right. \nonumber \\ 
	&+{\boldsymbol\Delta}^{(s_{2})}_{q_{2}}\expec{\textbf Q(\delta_1^{[q_1]},\delta_3^{[q_3]},\ldots,\delta_n^{[q_n]})}^{(s_1s_3\ldots s_n)} \nonumber \\ 
	&+\ldots \nonumber \\
	&\left. +{\boldsymbol\Delta}^{(s_{n})}_{q_{n}} \expec{\textbf Q(\delta_1^{[q_1]},\ldots,\delta_{n-1}^{[q_{n-1}]})}^{(s_1\ldots s_{n-1})} \right],  \label{eqn:NthSolOBE_2} 
	\end{align}
where, by convention, $\lim_{t\to \infty}\langle {\bf Q}(\ldots; t)\rangle^{(\ldots)}=\langle {\bf Q}(\ldots)\rangle^{(\ldots)}$. Comparing (\ref{eqn:NthSolOBE_2}) and (\ref{eqn:NthSolOBE}), we notice that the former is nothing but the recursive representation of the latter in terms of the $(n-1)$th-order corrections, provided that the $(n-1)$th-order corrections are given by Eq.~(\ref{eqn:N-1thSolOBE}). This finally proves that, if Eq.~(\ref{eqn:N-1thSolOBE}) is true, then also Eq.~(\ref{eqn:NthSolOBE}) is true. {\it Q. E. D.}

\subsection{Inelastic spectral responses}
\label{sec:inelastic}
\subsubsection{General solutions}
The inelastic part of the spectral responses arises from the fluctuating part of the stationary atomic dipole correlation function $\expec{\Delta D_q^{\dagger}(t)\Delta D_{r}(t^\prime)}$, where $\Delta D_r=D_{r}-\expec{D_{r}}$ ($\Delta D^{\dagger}_q=D^{\dagger}_{q}-\expec{D^{\dagger}_{q}}$), $q,r=\pm 1,0$, and $D^{\dagger}_q=\hat{\bf e}_q\cdot {\bf D}^\dagger$ ($D_{r}=\hat{\bf e}^*_r\cdot {\bf D}$) is the $q$th ($r$th) component of the atomic raising (lowering) operator (see Sec. \ref{sec:generalOBE}). To find this correlation function, we introduce two vectors:
 \begin{eqnarray}
     \textbf{f}_{r}(\tau)&=&\expec{\Delta \textbf{Q}(\tau)\Delta  D_{r}}, \label{eqn:FlucDipoleCorrelation1} \\
     \textbf{h}_{q}(\tau)&=&\expec{\Delta  D^\dagger_q \Delta \textbf{Q}(\tau)},
    \label{eqn:FlucDipoleCorrelation2}
    \end{eqnarray}
where $\tau=t^\prime-t$, if $t^\prime\geq t$, and $\tau=t-t^\prime$ otherwise. According to the quantum regression theorem \cite{scully}, both $\textbf{f}_{r}(\tau)$ and $\textbf{h}_{q}(\tau)$ obey the same equation of motion, but with different initial conditions [to be specified in Sec.~\ref{sec:InitialCond}].  Omitting for brevity the temporal argument, we can write
\begin{equation}
\dot{\textbf{f}}_r={\bf M}\textbf f_r+(e^{is_1\delta_1 t}\boldsymbol{\Delta}_{q_1}^{(s_1)}\textbf f_r+\ldots
+e^{is_n\delta_n t}\boldsymbol{\Delta}_{q_n}^{(s_n)}\textbf f_r),\label{f_l}
\end{equation}
and $\dot{\textbf{h}}_q$ is equal to the right hand side of (\ref{f_l}), with the ${\bf f}_r$ replaced by the ${\bf h}_q$.
Equation (\ref{f_l})  coincides with the generalized OBE, Eq. (\ref{Q_n}), up to a constant vector ${\bf L}$. As a result, both vectors, ${\bf f}_r$ and ${\bf h}_q$, tend to zero in the long-time limit (that is, the temporal correlations vanish).

As shown in \cite{Kneddi}, the $n$th-order correction ($0\leq n\leq N$) to the Laplace transform solution $\tilde{\bf f}_r(z^{\prime\prime})=\lim_{z^\prime\to 0} \tilde{\bf f}_r(z)$ of Eq. (\ref{f_l}) reads:
\begin{widetext}
\begin{align}
\tilde{\textbf f}_{r}^{(s_1\ldots s_n)}&(\delta_1^{[q_1]},\ldots,\delta_n^{[q_n]};z^{\prime\prime}) \nonumber \\
&=\sum_{\pi(j_1,\ldots,j_n)} \left[{\bf G}(iz^{\prime\prime}+i\zeta)\boldsymbol{\Delta}_{q_{j_n}}^{(s_{j_n})}\ldots \boldsymbol{\Delta}_{q_{j_2}}^{(s_{j_2})} {\bf G}(i z^{\prime\prime}+i s_{j_1} \delta_{j_1})\boldsymbol{\Delta}_{q_{j_1}}^{(s_{j_1})} {\bf G}(iz^{\prime\prime}) {\bf f}_r^{(0)}(0)  \right] \nonumber \\
        &+\sum_{\pi(j_2,\ldots,j_n)} \left[{\bf G}(iz^{\prime\prime}+i\zeta)\boldsymbol{\Delta}_{q_{j_n}}^{(s_{j_n})}\ldots \boldsymbol{\Delta}_{q_{j_2}}^{(s_{j_2})} {\bf G}(iz^{\prime\prime}+i s_{j_1} \delta_{j_1}){\bf f}_r^{(s_{j_1})}(\delta_{j_1}^{[q_{j_1}]};0)  \right] \nonumber \\
&+\ldots \nonumber \\
&+ \textbf {\bf G}(iz^{\prime\prime}+i\zeta) \textbf f_{r}^{(s_1\ldots s_n)}(\delta_1^{[q_1]},\ldots,\delta_n^{[q_n]};0),   \label{eqn:Laplacef}
\end{align}
\end{widetext}
where the $\textbf f_{r}^{(s_1...s_n)}(\delta_1^{[q_1]},...,\delta_n^{[q_n]};0)$ denote the initial conditions of corresponding order in the perturbative expansion of $\textbf f_r$ [see Sec.~\ref{sec:InitialCond}]. The solutions $\tilde{\textbf h}_{q}^{(s_1\ldots s_n)}(\delta_1^{[q_1]},\ldots,\delta_n^{[q_n]};z^{\prime\prime})$ follow after replacements, in the right hand side of Eq.~(\ref{eqn:Laplacef}), of the initial conditions $\textbf h_{q}^{(s_1...s_n)}(\delta_1^{[q_1]},...,\delta_n^{[q_n]};0)$ [see Sec.~\ref{sec:InitialCond}].
We skip the proof of Eq. (\ref{eqn:Laplacef}), which can be done inductively in the same way as the proof of Eq. (\ref{eqn:NthSolOBE}) (see Sec. \ref{app:ProofOBV}). Finally, in full analogy with the case of the elastic building blocks (see Sec. \ref{app:gen_sol}), the inelastic building blocks are obtained by taking certain elements of the  vectors $\tilde{\textbf f}_{r}^{(s_1\ldots s_n)}(\delta_1^{[q_1]},\ldots,\delta_n^{[q_n]};z^{\prime\prime})$ and $\tilde{\textbf h}_{q}^{(s_1\ldots s_n)}(\delta_1^{[q_1]},\ldots,\delta_n^{[q_n]};z^{\prime\prime})$ \cite{ralf13}. 

\subsubsection{Initial conditions}\label{sec:InitialCond}
The initial conditions which enter the Laplace transform solutions (\ref{eqn:Laplacef}), can be calculated from the perturbative solutions of the generalized polychromatic OBE (\ref{eqn:NthSolOBE}). To see this, we write Eq.~(\ref{eqn:FlucDipoleCorrelation1}) and (\ref{eqn:FlucDipoleCorrelation2}) in the following form:
\begin{eqnarray}
     \textbf{f}_{r}(0)&=&\expec{\textbf{Q}D_{r}}-\expec{\textbf{Q}}\expec{D_{r}}, \label{eqn:FlucDipoleCorrelationInitial1} \\
     \textbf{h}_{q}(0)&=&\expec{D^\dagger_q \textbf{Q}}-\expec{D^\dagger_q }\expec{\textbf{Q}},
    \label{eqn:FlucDipoleCorrelationInitial2}
\end{eqnarray}
The non-factorized parts on the right-hand side of Eqs.~(\ref{eqn:FlucDipoleCorrelationInitial1}) and (\ref{eqn:FlucDipoleCorrelationInitial2}) can be represented as follows:
\begin{eqnarray}
     \expec{\textbf{Q}D_{r}}=\textbf A_1 \expec{\textbf Q} +\textbf L_1, \label{eqn:NonFactorizingInitial1} \\
     \expec{D^\dagger_q \textbf{Q}}=\textbf A_2 \expec{\textbf Q} +\textbf L_2,
    \label{eqn:NonFactorizingInitial2}
\end{eqnarray}
with the matrices
 \begin{align}
    (\textbf A_1)_{ij}&=\frac{1}{4} \text{tr}[\mu_i D_r \mu^{\dagger}_j], \hspace{0.5cm} (L_1)_{i}=\frac{1}{4} \text{tr}[\mu_i^{\dagger} D_r ], \label{eqn:matricesA_12a} \\
    (\textbf A_2)_{ij}&=\frac{1}{4} \text{tr}[D^{\dagger}_q \mu_i  \mu^{\dagger}_j], \hspace{0.5cm} (L_2)_{i}=\frac{1}{4} \text{tr}[\mu_i^{\dagger}  D^{\dagger}_q ].
    \label{eqn:matricesA_12b}
    \end{align}
Next, we perform a perturbative expansion of both sides of Eqs.~(\ref{eqn:FlucDipoleCorrelationInitial1}) and (\ref{eqn:FlucDipoleCorrelationInitial2}). This yields:
\begin{widetext}
\begin{align}
    \textbf{f}_{r}^{(s_1...s_n)}(\delta_1^{[q_1]},...,\delta_n^{[q_n]};0)=&\textbf A_1 \expec{\textbf{Q}(\delta_1^{[q_1]},...,\delta_n^{[q_n]})}^{(s_1...s_n)}  \nonumber \\    &-\left(\expec{\textbf{Q}(\delta_1^{[q_1]},...,\delta_n^{[q_n]})}^{(s_1...s_n)} \expec{D_r}^{(0)} \right. \nonumber \\
    &+\sum_{\pi(j_1,...,j_{n-1}|j_n)} \expec{\textbf{Q}(\delta_1^{[q_1]},...,\delta_{n-1}^{[q_{n-1}]})}^{(s_{j_{1}}...,s_{j_{n-1}})} \expec{D_r(\delta_{j_n})}^{(s_{j_N})} \nonumber \\
    &+... \nonumber \\
    &+\sum_{\pi(j_1|j_2,...,j_n)}  \expec{\textbf{Q}(\delta_{j_1})}^{(s_{j_{1}})}  \expec{D_r(\delta_2^{[q_2]},...,\delta_n^{[q_n]})}^{(s_{j_{2}}...s_{j_{n}})} \nonumber \\
    &\left.+\expec{\textbf{Q}}^{(0)} \expec{D_r(\delta_1^{[q_1]},...,\delta_n^{[q_n]})}^{(s_1...s_n)}\right),   \label{eqn:Initial_f}
    \end{align}
\end{widetext}
where $\pi (j_1,...,j_k|j_{k+1},...,j_n)$ denotes the $n!/k!(n-k)!$ permutations between the two sets of indices $\{j_1,...,j_k\}$ and $\{j_{k+1},...,j_n\}$. In the simplest case of no incoming probe fields, Eq.~(\ref{eqn:Initial_f}) reduces to:
\begin{align}
    \textbf{f}_{r}^{(0)}(0)&=\textbf A_1 \expec{\textbf{Q}}^{(0)}+\textbf L_1-\expec{\textbf{Q}}^{(0)} \expec{D_r}^{(0)}.\label{eqn:0thq_12_1}
        \end{align}
    The expressions for $\textbf{h}_{q}^{(s_1...s_n)}(\delta_1^{[q_1]},...,\delta_N^{[q_n]};0)$ can be obtained by analogy to Eq.~(\ref{eqn:Initial_f}) and (\ref{eqn:0thq_12_1}), after the replacements $\textbf A_1 \rightarrow \textbf A_2$, $\textbf L_1\rightarrow \textbf L_2$, and $D_r \rightarrow D_q^{\dagger}$.

\bibliographystyle{prsty} 
\bibliography{mybib-1}

\begin{thebibliography}{10}

\bibitem{PhysRevLett.83.5266}
G. Labeyrie {\it et~al.}, Phys. Rev. Lett. {\bf 83},  5266  (1999).

\bibitem{PhysRevLett.88.203902}
Y. Bidel {\it et~al.}, Phys. Rev. Lett. {\bf 88},  203902  (2002).

\bibitem{akkermansMeso}
E. Akkermans and G. Montambaux, {\em Mesoscopic Physics of Electrons and
  Photons} (Cambridge University Press, New York, 2007).

\bibitem{PhysRevE.70.036602}
T. Chaneli\`ere {\it et~al.}, Phys. Rev. E {\bf 70},  036602  (2004).

\bibitem{balik05}
S. Balik {\it et~al.}, Journ.\ Mod.\ Opt. {\bf 52},  2269  (2005).

\bibitem{kupriyanov06}
D.~V. Kupriyanov, I.~M. Sokolov, C.~I. Sukenik, and M.~D. Havey, Laser Phys.
  Lett. {\bf 3},  223  (2006).

\bibitem{mueller2011}
C. M\"uller and D. Delande,  in {\em Disorder and interference: localization
  phenomena}, {\em Les Houches 2009 - Session XCI: Ultracold Gases and Quantum
  Information}, edited by C. Miniatura (Oxford University Press, Oxford, 2011),
  Chap.~9.

\bibitem{akkermans08}
E. Akkermans, A. Gero, and R. Kaiser, Phys. Rev. Lett. {\bf 101},  103602
  (2008).

\bibitem{skipetrov14}
S.~E. Skipetrov and I.~M. Sokolov, Phys. Rev. Lett. {\bf 112},  023905  (2014).

\bibitem{savels07}
T. Savels, A.~P. Mosk, and A. Lagendijk, Phys. Rev. Lett. {\bf 98},  103601
  (2007).

\bibitem{guerin10}
W. Guerin {\it et~al.}, J.~Opt. {\bf 12},  024002  (2010).

\bibitem{Baudouin:2013yg}
Q. Baudouin {\it et~al.}, Nat Phys {\bf 9},  357  (2013).

\bibitem{PhysRevA.64.053804}
C.~A. M\"uller, T. Jonckheere, C. Miniatura, and D. Delande, Phys. Rev. A {\bf
  64},  053804  (2001).

\bibitem{0295-5075-61-3-327}
G. Labeyrie {\it et~al.}, Europhys. Lett. {\bf 61},  327  (2003).

\bibitem{kupriyanov03}
D.~V. Kupriyanov {\it et~al.}, Phys. Rev. A {\bf 67},  013814  (2003).

\bibitem{PhysRevA.73.013802}
T. Wellens, B. Gr\'emaud, D. Delande, and C. Miniatura, Phys. Rev. A {\bf 73},
  013802  (2006).

\bibitem{agarwal74}
G.~S. Agarwal, {\em Quantum Statistical Theories of Spontaneous Emission and
  their Relation to other Approaches} (Springer, Berlin, 1974).

\bibitem{scully}
M.~O. Scully and M.~S. Zubairy, {\em Quantum Optics} (Cambridge University
  Press, Cambridge, 1997).

\bibitem{cohen1998atom}
C. Cohen-Tannoudji, J. Dupont-Roc, and G. Grynberg, {\em Atom-Photon
  Interactions: Basic Processes and Applications} (WILEY-VCH Verlag, Weinheim,
  1998).

\bibitem{PhysRevLett.94.043603}
V. Shatokhin, C.~A. M\"uller, and A. Buchleitner, Phys. Rev. Lett. {\bf 94},
  043603  (2005).

\bibitem{PhysRevA.73.063813}
V. Shatokhin, C.~A. M\"uller, and A. Buchleitner, Phys. Rev. A {\bf 73},
  063813  (2006).

\bibitem{gremaud06}
B. Gr\'emaud, T. Wellens, D. Delande, and C. Miniatura, \pra {\bf 74},  033808
  (2006).

\bibitem{PhysRevA.76.043832}
V. Shatokhin, T. Wellens, B. Gr\'emaud, and A. Buchleitner, Phys. Rev. A {\bf
  76},  043832  (2007).

\bibitem{Geiger2010244}
T. Geiger, T. Wellens, V. Shatokhin, and A. Buchleitner, Photon. Nanostr. -
  Fund. Appl. {\bf 8},  244   (2010).

\bibitem{PhysRevA.82.013832}
T. Wellens, T. Geiger, V. Shatokhin, and A. Buchleitner, Phys. Rev. A {\bf 82},
   013832  (2010).

\bibitem{binninger12}
T. Binninger, Multiple scattering of intense laser light in a cloud of cold
  atoms, Diploma thesis, Albert-Ludwigs-Universit{\"a}t Freiburg), 2012,
  \url{http://www.freidok.uni-freiburg.de/volltexte/8812}.

\bibitem{Shatokhin2010150}
V. Shatokhin, T. Geiger, T. Wellens, and A. Buchleitner, Chem. Phys. {\bf 375},
   150   (2010).

\bibitem{SlavaTriple}
V. Shatokhin and T. Wellens, Phys. Rev. A {\bf 86},  043808  (2012).

\bibitem{ralf}
R. Blattmann, The pump-probe approach to coherent backscattering of intense
  laser light by cold atoms with degenerate energy levels, Diploma thesis,
  Albert-Ludwigs-Universit{\"a}t Freiburg, 2011,
  \url{http://www.freidok.uni-freiburg.de/volltexte/9185}.

\bibitem{ralf13}
V.~N. Shatokhin, R. Blattmann, T. Wellens, and A. Buchleitner, Phys. Rev. A
  {\bf 90},  023850  (2014).

\bibitem{Kneddi}
A. Ketterer, The diagrammatic pump-probe approach to coherent backscattering of
  intense laser light by cold Sr atoms, Diploma thesis,
  Albert-Ludwigs-Universit{\"a}t Freiburg, 2013,
  \url{http://www.freidok.uni-freiburg.de/volltexte/9157}.

\bibitem{SlavaDiagrams}
V. Shatokhin, T. Wellens, and A. Buchleitner, J. Phys. B {\bf 45},  215501
  (2012).

\bibitem{wellens04}
T. Wellens, B. Gr\'emaud, D. Delande, and C. Miniatura, Phys. Rev. A {\bf 70},
  023817  (2004).

\bibitem{shatokhin10a}
V.~N. Shatokhin and S.~Y. Kilin, Opt. Spectr. {\bf 108},  446  (2010).

\bibitem{Heiderich1994110}
A. Heiderich, A.~S. Martinez, R. Maynard, and B.~A. van Tiggelen, Physics
  Letters A {\bf 185},  110   (1994).

\end{thebibliography}

\end{document}